\documentclass{llncs}

\usepackage{times}
\usepackage{amsmath}

\usepackage{epsfig}

\newtheorem{Example}{Example}

\newtheorem{Proposition}{Proposition}

\newenvironment{Proof}
              {{\noindent{\bf Proof:}}\normalfont\slshape}
              {\hfill\rule{2mm}{2mm}\\}  

\title{On Solving Boolean Multilevel Optimization Problems}
\author{Josep Argelich~\inst{1} \and In\^es Lynce~\inst{2} \and Joao Marques-Silva~\inst{3}}

\institute{INESC-ID\\
Lisboa, Portugal\\
\texttt{josep@sat.inesc-id.pt}
\and
IST/INESC-ID\\
Lisboa, Portugal\\
\texttt{ines@sat.inesc-id.pt}
\and
University College Dublin\\
Dublin, Ireland\\
\texttt{jpms@ucd.ie}
}
\begin{document}

\maketitle

%
%
%
%

\begin{abstract}
Many combinatorial optimization problems entail a number of hierarchically dependent optimization problems. An often used solution is to associate a suitably large cost with each individual optimization problem, such that the solution of the resulting aggregated optimization problem solves the original set of hierarchically dependent optimization problems. This paper starts by studying the package upgradeability problem in software distributions. Straightforward solutions based on Maximum Satisfiability (MaxSAT) and pseudo-Boolean (PB) optimization are shown to be ineffective, and unlikely to scale for large problem instances. Afterwards, the package upgradeability problem is related to multilevel optimization. The paper then develops new algorithms for Boolean Multilevel Optimization (BMO) and highlights a large number of potential applications. The experimental results indicate that the proposed algorithms for BMO allow solving optimization problems that existing MaxSAT and PB solvers would otherwise be unable to solve.

\end{abstract}


\section{Introduction}

Many real problems require an optimal solution rather than any solution. Whereas 
decision problems require a yes/no answer, optimization problems require the best 
solution, thus differentiating the possible solutions. In practice, there must be 
a classification scheme to determine how one solution compares with the others. 
Such classification may be seen as a way of establishing preferences that express 
cost or satisfaction.

A special case of combinatorial optimization problems may require a set of optimization criteria
to be observed, for which is possible to define a hierarchy of importance. Suppose that
instead of requiring a balance between price, horsepower and fuel consumption for 
choosing a new car, you have made a clear hierarchy in your mind: you have a 
strict limit on how much you can afford, then you will not consider a car with less than  
150 horsepower and after that the less the fuel consumption the better. Not only you establish
a hierarchy in your preferences, but also the preferences are defined in such a way that the
set of potential solutions gets subsequently reduced.
%
Such kind of problems are present not only in your daily life but also in many real applications.

Clearly, the kind of problems we target can be encoded as a constraint optimization problem,
making use of the available technology for dealing with preferences. 
Preference handling is one of the current hot topics in AI with active research lines
in constraint satisfaction and optimization~\cite{rossi-aim08}. Broadly, preferences 
over constraints may be expressed quantitatively or qualitatively. For example, one may 
wish to fly in the afternoon or simply choose the less expensive flight of that day.
Soft constraints model quantitative preferences by associating a 
level of satisfaction with each of the solutions~\cite{meseguer-hcp06}, whereas CP-nets 
model qualitative preferences by expressing preferential dependencies with pairwise 
comparisons~\cite{boutilier-jair04}. 
Furthermore, preference-based search algorithms can be generalized 
to handle multi-criteria optimization~\cite{junker-aor04}.

A straightforward approach to solve a special case of a constraint optimization problem, 
for which there is a total ranking of the criteria, would be to establish a lexicographic 
ordering over variables and domains, such that optimal solutions would come first in the 
search tree~\cite{freuder-soft03}. 
But this has the potential disadvantage of producing a thrashing behavior whenever
assignments that are not supported by any solution are considered, as a result of 
decisions made at the first nodes of the search tree~\cite{junker-aor04}.

Maximum satisfiability (MaxSAT) naturally encodes a constraint optimization problem
over Boolean variables where constraints are encoded as clauses. A solution to
the MaxSAT problem maximizes the number of satisfied clauses. Weights may also be 
associated with clauses, in which case the sum of the weights of the satisfied clauses is to
be maximized. The use of the weighted MaxSAT formalism allows to solve a set of 
hierarchically dependent optimization problems. Pseudo-Boolean (PB) optimization
may also be used to solve this kind of problems, given that weighted MaxSAT problem instances 
can be translated to PB. Each clause is extended with a relaxation variable that 
is then included in the cost function, jointly with the respective weight.

Boolean satisfiability (SAT) and PB have been extended in the past to handle preferences.
For example, SAT-based planning has been extended to include 
conflicting preferences~\cite{giunchiglia-aaai07}, for each of which weights are 
associated, thus requiring the use of an objective function involving the 
preferences and their weights. The proposed solution modifies a SAT backtracking
algorithm to search first for optimal plans by branching according to the partial
order induced by the preferences. In addition, algorithms for dealing with 
multi-objective PB problems have been developed~\cite{lukasiewycz-sat07},
in contrast to traditional algorithms that optimize a single linear function.

This paper is organized as follows. The next section describes the problem of package
upgradeability in software systems. This problem comes from a real application
and has been the drive for the algorithms 
being developed. Section~\ref{sec:mo} introduces multilevel optimization and relates it with a
variety of problems. Afterwards, specific multilevel optimization algorithms are
proposed, being based in MaxSAT and PB. Experimental results show the effectiveness 
of the new algorithms. Finally, the paper concludes.

\section{A Practical  Example}

We have all been through a situation where the installation of a new piece of software
turns out to be a nightmare. Not only you do not get the new computer program installed, but
also some other programs may eventually stop working properly. And this may also happen
when you simply want to upgrade to a more recent version of a program that you have been 
using for some time.
Although this seems to be a software engineering problem, behind the nightmare is a 
hard computational problem, and therefore an intelligent solution is desirable.

These kind of problems may occur because there are {\em constraints} 
between the different pieces of software (called {\em packages}).
Although these constraints are expected to be handled in a consistent and efficient way,
current software distributions are developed by distinct individuals. This is opposed to traditional 
systems which have a centralized and closed development. Open systems also tend to be much more 
complex, and therefore some packages may become incompatible. In such circumstances, user preferences
should be taken into account. For example, you would rather prefer to have your old version of {\em skype} 
working than to have the latest version not working properly.

The constraints associated with each package can be defined by a tuple $(p, D, C)$, where $p$ is the package, $D$ are the 
dependencies of $p$, and $C$ are the conflicts of $p$. $D$ is a set of dependency clauses, 
each dependency clause being a disjunction of packages. $C$ is a set of packages conflicting 
with $p$. 

Previous work has applied SAT-based tools to ensure the consistency of repositories and installations
as well as to upgrade consistently package installations. SAT-based tools have first been used
to support distribution editors~\cite{MBCVDLT06}. The developed tools are automatic and ensure completeness,
which makes them more reliable than ad-hoc and manual tools. Recently, Max-SAT has been
applied to solve the software package installation problem from the user point of view~\cite{AL08}. In addition, the
OPIUM tool~\cite{TSJL07} uses PB constraints and optimizes a user provided {\em single} objective function.
One modelling example could be preferring smaller packages to larger ones.

The encoding of these constraints into SAT is straightforward: for each package $p_i$
there is a Boolean variable $x_i$ that is assigned to true iff package $p_i$ is installed, and 
clauses are either dependency clauses or conflict clauses (one clause 
for each pair of conflicting packages).

\begin{Example}\label{ex-pc}
Given a set of package constraints $S = \{(p_1,\{ p_2, p_5 \lor p_6\}, \emptyset), (p_2, \emptyset, \{ p_3\}),$\\*
$ (p_3, \{ p_4\}, \{ p_1\}), (p_4, \emptyset,\{p_5,p_6\}) \}$, its encoded CNF instance is the following:
$$
\begin{array}{cc}
\neg x_1 \lor x_2 & \neg x_3 \lor x_4\\
\neg x_1 \lor x_5 \lor x_6 & \neg x_3 \lor \neg x_1\\
\neg x_2 \lor \neg x_3 & \neg x_4 \lor \neg x_5\\
& \neg x_4 \lor \neg x_6
\end{array}
$$
\end{Example}

The problem described above is called software {\em installability} problem.
The possibility of upgrading some of the packages (or introducing new packages)
poses new challenges as existing packages may eventually be deleted.
The goal of the software {\em upgradeability} problem is to find a solution that satisfies
user preferences by minimizing the impact of introducing new packages in the current system, 
which is a reasonable assumption.
Such preferences may be distinguished establishing the following hierarchy:
(1) constraints on packages cannot be violated, 
(2) required packages should be installed,
(3) packages that have been previously installed by the user should not be deleted,
(4) the number of remaining packages installed (as a result of dependencies) should be minimized.

The software upgradeability problem can be naturally encoded as
a weighted partial MaxSAT problem. In weighted MaxSAT, each clause is a pair $(C, w)$
where $C$ is a CNF clause and $w$ is its corresponding weight. In weighted partial
MaxSAT, {\em hard} clauses {\em must} be satisfied, in contrast to the remaining {\em soft}
clauses that {\em should} be satisfied. Hard clauses are associated with a weight that
is greater than the sum of the weights of the soft clauses. A solution to the weighted partial 
MaxSAT problem maximizes the sum of the weights of the satisfied clauses. However, if the 
solution found exceeds the weight associated with a hard clause, then at least one hard clause
is unsatisfied, and consequently the solver returns no solution.

The following example shows a weighted partial MaxSAT formula for the upgradeability problem.

\begin{Example}\label{ex-up}
Given a set of package constraints $S = \{(p_1,\{ p_2, p_5\}, \{ p_4\}), (p_2,
\emptyset, \emptyset),$\\*
$ (p_3, \{ p_2 \vee p_4\}, \emptyset), (p_4, \emptyset, \emptyset), (p_5, \emptyset,
\emptyset)\}$, the set of packages that the user wants to install $I = \{p_1\}$, and
the current set of installed packages in the system $A = \{p_2\}$, 
its encoded weighted partial MaxSAT instance is the following:
$$
\begin{array}{c}
(\neg x_1 \lor x_2, 16)\\
(\neg x_1 \lor x_5, 16)\\
(\neg x_1 \lor \neg x_4, 16)\\
(\neg x_3 \lor x_2 \lor x_4, 16)\\
(x_1, 8)\\
(x_2, 4)\\
(\neg x_3, 1)\\
(\neg x_4, 1)\\
(\neg x_5, 1)\\
\end{array}
$$
\end{Example}

In Example~\ref{ex-up}, clause weights depend on
the kind of clauses we are encoding.
We assign a {\em hard} weight (with value 16) to clauses encoding
the dependencies and conflicts.
A maximum weight (with value 8) is assigned to the packages the user 
wants to install.
A medium weight (with value 4) is assigned to clauses encoding
packages currently installed in our system in order to minimize the number of
removed packages. 
Finally, the minimum weight (with value 1) is assigned to
clauses encoding the remaining packages in order to minimize the number of
additional packages being installed as a result of dependencies.

As we can see, this weight distribution gives priority to the user preferences
over all the other packages, and it also gives priority to the current
installation profile over the remaining packages.


\section{Multilevel Optimization}
\label{sec:mo}

The software upgradeability problem described in the previous section
can be viewed as a special case of the more general problem of {\em
  Multilevel Optimization}~\cite{savard-aor07}\footnote{This problem
  is also referred to as {\em Multilevel
    Programming}~\cite{candler-tr77} and {\em Hierarchical
    Optimization}~\cite{anandalingam-aor92}.}. Multilevel optimization
can be traced back to the early 70s~\cite{bracken-or73}, when
researchers focused on mathematical programs with optimization
problems in the constraints. Multilevel optimization represents a
hierarchy of optimization problems, where the outer optimization
problem is subject to the outcome of each of the enclosed optimization
problems in order.
In part motivated by the practical complexity of the multilevel
optimization, most work in the recent past has addressed the special
case of bilevel optimization~\cite{savard-aor07}.
Moreover, and for the special case of integer or Boolean variables,
existing work is still preliminary~\cite{ralphs-informs09}.
It should also be observed that the general problems of bilevel and
multilevel optimization find a wide range of
applications~\cite{savard-aor07}, representative examples of which can
be represented with integer or Boolean variables~\cite{savard-orl04}.

One can conclude that the software upgradeability problem can be
viewed as an example of multilevel programming, where the constraints
are clauses, and the variables have Boolean domain. The least
constrained (or outer) optimization problem represents the problem of
minimizing the number of newly installed packages due to dependencies,
whereas the most constrained (or inner) optimization problem
represents the problem of maximizing the installation of packages in
the user preferences.

This paper focuses on the special case of multilevel optimization
where the constraints are propositional clauses and the variables have
Boolean domain. This problem will be referred to as {\em Boolean
  Multilevel Optimization}~(BMO). The hierarchy of optimization
problems can be captured by associating suitable weights with the
clauses, as illustrated for the package upgradeability problem.

More formally, consider a set of clauses $C = C_1\cup C_2\cup \cdots
\cup C_m$, where $C_1, C_2, \ldots,$ $C_m$ form a partition of
$C$. Moreover, consider the partition of $C$ as a sequence of sets of
clauses:
\begin{equation}
\label{eq:clset}
\langle C_1, C_2, \ldots, C_m\rangle
\end{equation}
Where a weight is associated with each set of clauses:
\begin{equation}
\label{eq:wset}
\langle w_1, w_2, \ldots, w_m\rangle
\end{equation}
As with MaxSAT, clauses with weight $w_m$ are required to be
satisfied, and so are referred to as {\em hard} clauses.
The associated optimization problem is to satisfy clauses in
$C_1\cup C_2\cup \ldots \cup C_m$ such that the sum of the weights of
the satisfied clauses is {\em maximized}.

Moreover, the hierarchy of optimization problems is captured by the
condition:
\begin{equation}
\label{eq:mbo}
w_i> \sum_{1\le j<i} w_j\cdot|C_j| \qquad i=2, \ldots, m
\end{equation}
The above condition ensures that the solution to the BMO problem can
be split into a sequence of optimization problems, first solving the
optimization problems for the soft clauses with the largest weight
(i.e.~$w_{m-1}$), then for the next clause weight, and so on until all
clause weights are considered.
Building on this observation, the next section proposes dedicated
algorithms for BMO.

\section{Solving Boolean Multilevel Optimization}

This section describes alternative solutions for BMO, in addition to
the weight-based solution described earlier in the paper. The first
solution is based on iteratively rescaling the weights of the MaxSAT
formulation. The second formulation extends the standard encoding of
weighted MaxSAT with PB constraints.

\subsection{BMO with MaxSAT}

Consider the BMO problem specified by equations (\ref{eq:clset}),
(\ref{eq:wset}) and (\ref{eq:mbo}).
%

The use of MaxSAT considers a sequence of $m - 1$ MaxSAT problems, at
each step rescaling the weights of the clauses and the initial upper
bound (UB) for each problem.
The first MaxSAT problem is defined as follows:

\begin{equation}
\label{eq:ms-base}
\begin{array}{l}
\textrm{Initial UB: } |C_{m-1}| + 1 \\[7pt]
\bigwedge_{c\in C_m} (c, |C_{m-1}| + 1)\\[3pt]
\bigwedge_{c\in C_{m-1}} (c, 1)\\
\end{array}
\end{equation}

Let the optimum solution be $u_{m-1}$, that represents the {\em
minimum} sum of weights of falsified clauses\footnote{The MaxSAT problem
is often referred as the MinUNSAT problem.}. In this case, as the weights
of clauses to optimize is one, $u_{m-1}$ will be the minimum number
of falsified clauses. The remaining MaxSAT problems can then be
defined as follows:

\begin{equation}
\label{eq:ms-step}
\begin{array}{l}
\textrm{Initial UB: } (u_{m-1}+1)\cdot p_{m-1} \\[7pt]
\bigwedge_{j=i+2}^{m} \wedge_{c\in C_j} (c, (u_{j-1}+1)\cdot p_{j-1}) \\[3pt]
\bigwedge_{c\in C_{i+1}} (c, (|C_i|+1)\cdot p_i)\\[3pt]
\bigwedge_{c\in C_i} (c, 1)\\
\end{array}
\end{equation}
With $1\le i<m-1$, where $p_i$ is the weight used for the set of clauses
$C_i$ within the same subproblem, and the optimum solution is $u_i$.
Observe that the value of $p_{m-1}$ is refined for each iteration of
the algorithm.
%
Also note that in this case, $u_i$ represents the minimum number of
clauses that needs to be falsified for clause weight $w_i$,
taking into account that for larger weights, the number of falsified
clauses {\em must} be taken into account.
The last problem to be considered corresponds to $i = 1$, for the
clauses with the smallest weight.

Finally, the MaxSAT solution for the original problem is obtained as follows:
\begin{equation}
\label{eq:ms-opt}
\sum_{i=1}^{m-1} w_i \cdot (|C_i| - u_i)
\end{equation}

\begin{Proposition}
The value obtained with (\ref{eq:ms-opt}), where the different $u_i$
values are obtained by the solution of the (\ref{eq:ms-base}) and 
(\ref{eq:ms-step}) MaxSAT problems, yields the correct solution to the
BMO problem.
\end{Proposition}

\begin{Proof}(Sketch) Proof follows from the above explanation,
  taking into account the condition on clauses' weights
  (\ref{eq:mbo}).
\end{Proof}

\subsection{BMO with PB}

The efficacy of the rescaling method of the previous section is still
bound by the weights used. Even though the rescaling method is
effective at reducing the weights that need to be considered, for very
large problem instances the challenge of large clause weights can
still be an issue.
An alternative approach is described in this section, which eliminates
the need to handle large clauses weights. This approach is based on
solving the BMO problem as a sequence of PB problems.


Consider the BMO problem specified by equations (\ref{eq:clset}),
(\ref{eq:wset}) and (\ref{eq:mbo}).
Each set of clauses $C_i$ can be modified by adding a relaxation
variable to each clause. The resulting set of relaxed clauses is
$C_i^r$, and the set of relaxation variables used is denoted by $Y_i$.
For example, if $c_j\in C_i$, the resulting clause is
$c_{j,r} = c_j\cup y_j$, and $y_j\in Y_i$. 
The technique of solving MaxSAT by using relaxation variables to
clauses is a standard technique~\cite{dlb-ictai96,kas-iccad02}.

The next step is to solve a sequence of PB problems. The first PB
problem is defined as:
\begin{equation}
\label{eq:pb-base}
\begin{array}{l}
\textrm{min} \qquad \sum_{y\in Y_{m-1}} \;y \\[5pt]
\textrm{s.t.} \qquad
\begin{array}[t]{l}
\bigwedge_{c\in C_m} \;c \\[3pt]
\bigwedge_{c_r\in C_{m-1}^r} \;c_r\\
\end{array}\\
\end{array}
\end{equation}
Let the optimum solution be $v_{m-1}$. $v_{m-1}$ represents the {\em
  largest} number of clauses that can be satisfied, independently of
the other clause weights.

Moreover, the remaining PB problems can then be defined as follows:
\begin{equation}
\label{eq:pb-step}
\begin{array}{l}
\textrm{min} \qquad \sum_{y\in Y_{i}} \;y \\[5pt]
\textrm{s.t.} \qquad
\begin{array}[t]{l}
\bigwedge_{c\in C_m} \;c \\[3pt]
\bigwedge_{j=i}^{m-1} \left(\wedge_{c_r\in C_{j}^r} c_r \right)\\[3pt]
\bigwedge_{j=i+1}^{m-1}\left( \sum_{y\in Y_j} y = v_j \right)\\
\end{array}\\
\end{array}
\end{equation}
With $1\le i<m-1$, and where the optimum solution is $v_i$. In this
case, $v_i$ represents the largest number of clauses that can be
satisfied for clause weight $w_i$, taking into account that for larger
weights, the number of satisfied clauses {\em must} be taken into
account.
The last problem to be considered corresponds to $i = 1$, for the
clauses with the smallest weight.

Finally, given the definition of $v_i$, the PB-based BMO solution is
obtained as follows:
\begin{equation}
\label{eq:pb-opt}
\sum_{i=1}^{m-1} w_i\cdot v_i
\end{equation}

As can be concluded, the proposed PB-based approach can solve the BMO
problem without directly manipulating any clause weights.

\begin{Proposition}
The value obtained with (\ref{eq:pb-opt}), where the different $v_i$
values are obtained by the solution of the (\ref{eq:pb-base}) and
(\ref{eq:pb-step}) PB problems, yields the correct solution to the
BMO problem.
\end{Proposition}

\begin{Proof}(Sketch) Proof follows from the above explanation,
  taking into account the condition on clauses' weights
  (\ref{eq:mbo}).
\end{Proof}

\section{Experimental Evaluation}

This section describes the experimental evaluation conducted to show the effectiveness of the
new algorithms described above. With this purpose, we have generated a comprehensive set of problem 
instances of the software upgradeability problem. In a first step, a number of off-the-shelf MaxSAT 
and PB solvers have been run. In a second step, these MaxSAT and PB solvers have been adapted to 
perform BMO approaches. In what follows we will use BMO$^{rsc}$ to denote weight rescaling BMO with MaxSAT
and BMO$^{ipb}$ to denote BMO with iterative pseudo-Boolean optimization.

\subsection{Experimental Setup}

The problem instances of the upgradeability problem have been obtained from the Linux Debian 
distribution archive\footnote{\tt http://snapshot.debian.net}, where Debian packages are daily archived. 
Each daily archive is a repository. Two repositories corresponding to a snapshot with a time gap of 6 
months have been selected. From the first repository, the packages
for a basic Debian installation have been picked, jointly with a set of other packages. From the second
repository, a set of packages to be upgraded have been picked. This set of packages is a subset of the 
installed packages. 

Each problem instance is denoted as {\tt i<x>u<y>} where {\tt x} is the number of installed packages 
(apart from the 826 packages of the basic installation) and {\tt y} is the number of packages to be 
upgraded. In the following experiments the number of {\tt x} packages ranges from 0 to 2000 and the number 
of {\tt y} packages is 98. The {\tt y} packages correspond to the subset of packages of the basic installation 
that have been updated from one repository to the other.

The four MaxSAT solvers\footnote{\tt http://www.maxsat.udl.cat/} used for the evaluation are:
IncWMaxSatz~\cite{LSL08}, MiniMaxSat~\cite{HLO08}, Sat4jMaxsat~\cite{sat4j} and WMaxSatz~\cite{AM07}.
The four PB solvers\footnote{\tt http://www.cril.univ-artois.fr/PB07/} used for the evaluation are:
Bsolo~\cite{MM04}, Minisat+~\cite{ES06}, PBS4~\cite{ARMS02} and Sat4jPB~\cite{sat4j}.
Other solvers could have been used, even tough we believe that these ones are some of the most 
competitive and overall implement different techniques which affect performance differently.
For each solver a set of instances were run with the default solver and BMO$^{rsc}$
or BMO$^{ipb}$.

Furthermore, for the best performing solver an additional number of instances has been run in order to study  
the scalability of the solver as the number of packages to install increases.

The experiments were performed on an Intel Xeon 5160 server (3.0GHz, 1333Mhz FSB, 4MB cache) running 
Red Hat Enterprise Linux WS 4. For Sat4j was used JRE 1.6.0\_0.07. For each instance was given the 
timeout of 900 seconds.

\subsection{Experimental Results}

Table~\ref{tab-sat} shows the CPU time required by MaxSAT solvers to solve a set of given problem instances. Column
{\em Default} shows the results for the off-the-shelf solver and column BMO$^{rsc}$ shows the results for 
the weight rescaling approach specially designed for solving BMO problems with MaxSAT.
For each instance the best result is highlighted in bold.

Clearly, IncWMaxSatz with BMO$^{rsc}$ is the best performing solver. Nonetheless, every other solver
benefits from the use of BMO$^{rsc}$. The only exception is Sat4jMaxsat because it spends around 8 seconds 
to read each instance and with BMO$^{rsc}$ the solver is called three times for the instances considered. Another advantage of 
using BMO$^{rsc}$ is that the solvers do not need to deal with the large integers representing the clause weights, which are used in the default encoding. 
This can be such a serious issue that for some solvers there are a few problem instances (designated with '-')
that cannot be solved.

\begin{table*}[ht]
\footnotesize
\begin{center}
\begin{tabular}{|l||r|r|r|r|r|r|r|r|}
\hline
 & \multicolumn{2}{|c|}{IncWMaxSatz} & \multicolumn{2}{|c|}{MiniMaxSat} & \multicolumn{2}{|c|}{Sat4jMaxsat} & \multicolumn{2}{|c|}{WMaxSatz} \\ \hline
Instance & \multicolumn{1}{|c|}{Default} & \multicolumn{1}{|c|}{BMO$^{rsc}$} & \multicolumn{1}{|c|}{Default} & \multicolumn{1}{|c|}{BMO$^{rsc}$} & \multicolumn{1}{|c|}{Default} & \multicolumn{1}{|c|}{BMO$^{rsc}$} & \multicolumn{1}{|c|}{Default} & \multicolumn{1}{|c|}{BMO$^{rsc}$} \\ \hline
i0u98     &  3.90           &  \textbf{3.29}  &  -      &  89.96 &  10.74  &  29.78  &  275.50  &  13.15 \\ \hline
i10u98    &  \textbf{3.58}  &          3.63   &  -      &  90.06 &  10.60  &  25.88  &  276.32  &  13.19 \\ \hline
i20u98    &  4.72           &  \textbf{3.67}  &  -      &  90.24 &  10.77  &  25.94  &  348.13  &  13.28 \\ \hline
i30u98    &  4.33           &  \textbf{3.81}  &  -      &  90.39 &  10.80  &  26.02  &  316.93  &  14.87 \\ \hline
i40u98    &  4.13           &  \textbf{3.58}  &  254.21 &  92.20 &  10.37  &  26.67  &  265.45  &  14.67 \\ \hline
i50u98    &  4.57           &  \textbf{3.37}  &  -      &  91.65 &  -      &  27.53  &  -       &  18.67 \\ \hline
i100u98   &  7.50           &  \textbf{3.97}  &  -      &  99.79 &  -      &  26.54  &  -       & 100.98 \\ \hline
i200u98   & 16.22           &  \textbf{5.64}  &  -      &  95.89 &  -      &  27.57  &  -       & $>$900 \\ \hline
i500u98   & 22.98           &  \textbf{4.82}  &  -      & 126.97 &  -      &  46.51  &  -       & $>$900 \\ \hline
i1000u98  & 37.47           &  \textbf{5.74}  &  -      & 195.54 &  -      & $>$900  &  -       & $>$900 \\ \hline
i2000u98  & 45.69           &  \textbf{7.39}  &  -      & 223.81 &  -      & 685.17  &  -       & $>$900 \\ \hline
\end{tabular}                 
\end{center}                  
\caption{The software upgradeability problem with weighted partial MaxSAT solvers (time in seconds)}
\label{tab-sat}               
\end{table*}                  

\begin{table*}[ht]
\footnotesize
\begin{center}
\begin{tabular}{|l||r|r|r|r|r|r|r|r|}
\hline
 & \multicolumn{2}{|c|}{Bsolo} & \multicolumn{2}{|c|}{Minisat+} & \multicolumn{2}{|c|}{PBS4} & \multicolumn{2}{|c|}{Sat4jPB} \\ \hline
Instance & \multicolumn{1}{|c|}{Default} & \multicolumn{1}{|c|}{BMO$^{ipb}$} & \multicolumn{1}{|c|}{Default} & \multicolumn{1}{|c|}{BMO$^{ipb}$} & \multicolumn{1}{|c|}{Default} & \multicolumn{1}{|c|}{BMO$^{ipb}$} & \multicolumn{1}{|c|}{Default} & \multicolumn{1}{|c|}{BMO$^{ipb}$} \\ \hline
i0u98     & 5.38   & 23.81  & $>$900  & 5.97  & $>$900  & 116.45  & {\bf 3.97}  & 11.72  \\ \hline 
i10u98    & 25.33   & 23.63    & $>$900  & 5.91  & $>$900  & 46.26  & {\bf 3.63}  & 11.67  \\ \hline 
i20u98    & 91.13   & 23.37    & $>$900  & {\bf 7.77}  & 735.54  & 59.11  & 18.05  & 13.82  \\ \hline 
i30u98    & 104.18   & 23.25    & $>$900  & {\bf 7.83}  & $>$900  & 78.88  & 19.10  & 13.74  \\ \hline 
i40u98    & 92.27   & 23.13    & $>$900  & {\bf 22.52}  & $>$900  & 111.40  & 48.42  & 26.48  \\ \hline 
i50u98    & 103.73   & {\bf 23.00}    & $>$900  & 25.91  & $>$900  & 64.49  & 48.35  & 25.98  \\ \hline 
i100u98   & 321.46   & 22.40    & $>$900  & {\bf 19.22}  & $>$900  & 78.81  & 41.09  & 54.86  \\ \hline 
i200u98   & $>$900  & {\bf 22.19}    & $>$900  & 39.78  & $>$900  & 70.86  & 69.53  & 116.05  \\ \hline 
i500u98   & $>$900  & {\bf 23.61}    & $>$900  & $>$900  & $>$900  & 91.17  & 158.52  & $>$900  \\ \hline 
i1000u98  & $>$900  & {\bf 71.51}    & $>$900  & $>$900  & $>$900  & $>$900  & $>$900  & $>$900  \\ \hline 
i2000u98  & $>$900  & 90.15    & $>$900  & $>$900  & $>$900  & 242.10  & $>$900  & {\bf 40.54}  \\ \hline 
\end{tabular}
\end{center}
\caption{The software upgradeability problem with pseudo-Boolean solvers (time in seconds)}
\label{tab-pb}
\end{table*}

Table~\ref{tab-pb} shows the results for PB solvers on solving the same instances. 
BMO$^{ipb}$ boosts the solvers performance, being Sat4jPB the only exception (for some instances it improves,
for some other it does not). For the remaining solvers, the improvements are
significant: most of the instances aborted by the default solver are now solved with BMO$^{ipb}$. 
Although there is no dominating solver, in contrast to what happens with IncWMaxSatz in the MaxSAT 
solvers, Bsolo is the only solver able to solve all the instances with BMO$^{ipb}$. Also, despite existing
a trend to increasing run times as the size of the instances increase, there are a few outliers. This also
contrasts with MaxSAT solvers, but is no surprise as additional variables can degrade the solvers 
performance in an unpredictable way.

\begin{figure}[!ht]
\begin{center}
\epsfig{file=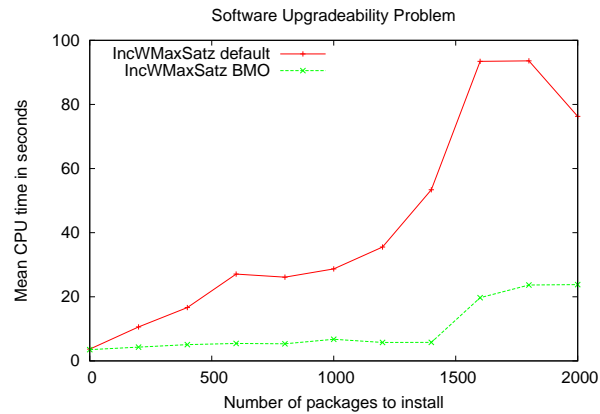, width=8cm}
\end{center}
\caption{Scalability of the solver IncWMaxSatz in its default version and using the BMO$^{rsc}$ when increasing the number of packages to install from 0 to 2000}
\label{fig-sat}
\end{figure}

Finally, we have further investigated IncWMaxSatz, which was the best performing solver. 
Figure~\ref{fig-sat} shows the scalability of the solver comparing the default performance of IncWMaxSatz 
with its performance using BMO$^{rsc}$. (The plot includes results for additional instances, with each point 
corresponding to the average of 100 instances.) We should first
note that the default IncWMaxSatz solver is by far more competitive than any other default MaxSAT or PB solvers.
Its performance is not even comparable with WMaxSatz, despite IncWMaxSatz being an extension of WMaxSatz. This 
is due to the features of IncWMaxSatz that make it particularly suitable for these 
instances, namely the incremental lower bound computation and the removal of inference rules that are particularly effective for solving random instances.
Nonetheless, BMO$^{rsc}$ has been able to improve its performance and to reduce the impact of the size of the 
instance in the performance.

\section{Conclusions}

In many practical applications, one often needs to solve a hierarchy
of optimization problems, where each optimization problem is
specified in terms of a sequence of nested optimization problems.
Examples in AI include specific optimization problems with
preferences.
Another concrete example is package management in software systems,
where SAT, PB and MaxSAT find increasing application.
It is possible to relate these optimization problems with multilevel
(or hierarchical)
optimization~\cite{bracken-or73,candler-tr77,savard-aor07}, which
finds a large number of practical applications. Despite the potential
practical applications, work on multilevel optimization algorithms
with linear constraints and integer or Boolean variables is still
preliminary~\cite{ralphs-informs09}.

This paper focus on Boolean Multilevel Optimization~(BMO) and, by
considering the concrete problem of package upgradeability in software
systems, shows that existing solutions based on either MaxSAT or PB
are in general inadequate.  Moreover, the paper proposes two different 
algorithms, one that uses MaxSAT and another that uses PB, to show
that dedicated algorithms for BMO can be orders of magnitude more
efficient than the best off-the-shelf MaxSAT and PB solvers.

Despite the very promising results, a number of research directions
can be outlined. One is to evaluate how the proposed algorithms scale
for larger problem instances. Another is to consider other
computational problems in AI that can be cast as BMO, for example in
the area of preferences and in the area of SAT-based optimization.

\bibliographystyle{abbrv}
\bibliography{main}

\end{document}